\documentclass[preprint]{elsarticle}
\usepackage{amssymb}
\usepackage[framed,numbered,autolinebreaks,useliterate]{mcode}
\usepackage{bm}
\usepackage{amsmath}
\usepackage{subfigure}
\usepackage{graphicx}
\usepackage{algorithm}
\usepackage{algorithmic}
\usepackage{mathrsfs}

\journal{Journal of Computational Physics}

\begin{document}

\begin{frontmatter}

\title{Reduction of Nonlinear Embedded Boundary Models for Problems with Evolving Interfaces}

\author[aa]{Maciej Balajewicz\corref{cor1}\fnref{fn1}}\ead{maciej.balajewicz@stanford.edu}
\author[aa,me,icme]{Charbel Farhat\fnref{fn2}}
\address[aa]{Department of Aeronautics and Astronautics}
\address[me]{Department of Mechanical Engineering}
\address[icme]{Institute for Computational and Mathematical Engineering \\ Stanford University, Stanford, CA 94305-4035, U.S.A}

\fntext[fn1]{Postdoctoral Fellow}
\fntext[fn2]{Vivian Church Hoff Professor of Aircraft Structures}
\cortext[cor1]{Corresponding author}

\begin{abstract}
Embedded boundary methods alleviate many computational challenges, including those associated with meshing complex geometries
and solving problems with evolving domains and interfaces. Developing model reduction methods for computational frameworks based
on such methods seems however to be challenging. Indeed, most popular model reduction techniques are projection-based,
and rely on basis functions obtained from the compression of simulation snapshots. In a traditional interface-fitted computational
framework, the computation of such basis functions is straightforward, primarily because the computational domain does not contain
in this case a fictitious region. This is not the case however for an embedded computational framework because the computational
domain typically contains in this case both real and ghost regions whose definitions complicate the collection and compression of
simulation snapshots. The problem is exacerbated when the interface separating both regions evolves in time. This paper addresses
this issue by formulating the snapshot compression problem as a weighted low-rank approximation problem where the
binary weighting identifies the evolving component of the individual simulation snapshots. The proposed approach is
application independent and therefore comprehensive. It is successfully demonstrated for the model reduction of several
two-dimensional, vortex-dominated, fluid-structure interaction problems.
\end{abstract}


\end{frontmatter}

\section{Introduction}
\label{sec:Introduction}
Two common computational frameworks for the solution of problems with evolving domains and interfaces
are the Arbitrary Lagrangian Eulerian (ALE)~\cite{Donea,CF1,CF2,ALERezoning} and Embedded Boundary Method (EBM)
frameworks. In an ALE framework, the computations are performed on an interface-fitted mesh which is deformed using
a mesh motion algorithm~\cite{MM1,MM2,MM3} to follow the evolution of the interface. In an EBM framework,
the interface is embedded in a background mesh --- also called an embedding mesh --- and allowed to intersect it as it
evolves~\citep{Peskin_AN_2002,Mittal_ARFM_2005,Wang_IJNMF_2011,FIVER}. In the latter case, the interface separates the
computational domain in two regions: a ``ghost'' (or fictitious) region usually associated with the volume delimited
by the embedded interface (boundary, or body), and a ``real'' region usually associated with the remainder of the
computational domain.  Both computational frameworks have advantages and shortcomings. An ALE framework is particularly
advantageous for
problems where the evolution of the interface is characterized by small-amplitude motions and deformations. An EBM framework
is often indispensable for problems where the interface undergoes large-amplitude motions and deformations, and/or topological
changes.

Irrespectively of the chosen framework however, the computational cost of high-fidelity simulations of problems with evolving
domains and interfaces can be prohibitive. For this reason, interest in Model Order Reduction (MOR) methods continues to
grow~\citep{Rewienski_LAA_2006,Thuan1,Amsallem1,Astrid_AC_2008,Chaturantabut_JSC_2010,Carlberg1,Balajewicz_JFM_2013,Carlberg_JCP_2013,Navon1,Navon2}. 
Most of these methods are projection-based. Therefore, they rely on the computation of suitable basis functions. Typically,
these are constructed by computing solution snapshots of some instances of the family of problems of interest and compressing
them using the Proper Orthogonal Decomposition (POD) or Singular Value Decomposition (SVD) method. In an interface-fitted
framework, this computation is straightforward because the computational domain does not contain a ghost region. In an
EBM framework however, this computation is complicated by the presence of a ghost region where the solution snapshots may or may
not be populated, and in the latter case, the populated values may or may not have a physical meaning. The situation is
further exacerbated when the interface evolves in time, as in this case the partitioning into meaningful and meaningless
components of the collected snapshots also evolves in time, which complicates further the compression of the collected snapshots.

Therefore, the main objective of this paper is to propose an effective method for the compression of solution snaphsots collected
during the simulation of problems with evolving domains and interfaces using an EBM, in view of enabling the projection-based
reduction of high-fidelity embedded boundary computational models.  This method consists of formulating the data compression
problem as a weighted low-rank matrix approximation problem, where the weighting is binary and identifies the ghost/real
partition of the individual snapshots. Therefore, the basis functions derived using this approach are optimal for the real
component of the partition of the computational domain.

The weighted low-rank approximation problem formulated in this paper is pervasive. It can be found, for example, in:
\begin{itemize}
\item Computer vision applications with occlusion.
\item Signal processing applications with irregular measurements in time and space.
\item Control problems with malfunctioning measuring devices.
\end{itemize}
This problem is a generalization of the popular compressed sensing problem~\citep{Donoho_2006,Candes_2008}. It also shares many
similarities with the matrix completion problem~\citep{Candes_2009,Cai_2010,Jain_ACM_2013,Vandereycken_SIAM_2013}.
For all these reasons, many algorithms for the solution of this problem are readily available (for example,
see~\citep{Srebro_ICML_2003,Buchanan_2005,Chen_2008}).

Whereas the approach outlined above for compressing solution snaphots computed by an EBM in view of constructing a reduced-order
basis suitable for model reduction is application independent and therefore quite general, it is developed and demonstrated in
the remainder for this paper for Fluid-Structure Interaction (FSI) problems. To this effect, the remainder of this paper is organized
as follows.

In Section~\ref{sec:NOT}, the main notation used in this paper is laid out to facilitate its reading. In Section~\ref{sec:EBM},
a general EBM framework for Computational Fluid Dynamics (CFD) is overviewed, and the standard projection-based MOR
approach is recapitulated. In Section \ref{sec:MET}, the proposed method for compressing CFD snapshots computed by an EBM
is introduced. In Section~\ref{sec:APP}, this method is applied to three different FSI problems. Finally in
Section~\ref{sec:CON}, conclusions are offered and prospects for future work are summarized.

\section{Notations and definitions}
\label{sec:NOT}

Throughout this paper, matrices are denoted by bold capitals (ex. $\bm{A}$), vectors by bold lower case (ex. $\bm{a}$),
and subscripts identify rows and columns (for example, $A_{i,j}$ is the entry at the $i$-th row and $j$-th column of
matrix $\bm{A}$). The ``colon'' notation is used to specify columns or rows of a matrix (for example $A_{:,j}$ denotes
the $j$-th column and $A_{1:n,:}$ denotes the first $n$ rows of matrix $\bm{A}$). $\bm{I}_{n}$ and $\bm{0}_{n}$ denote
the ${n \times n}$ identity and null matrices, respectively.

For two matrices $\bm{A}$ and $\bm{B}$ of equal dimension $m \times n$, the Hadamard product $\bm{A} \odot \bm{B}$ is
the matrix of the same dimension whose elements are given by
\begin{equation}
    \left( \bm{A} \odot \bm{B} \right)_{i,j}  = A_{i,j}  \cdot B_{i,j}
\end{equation}

If $\bm{A}$ is an $m \times n$ matrix and $\bm{B}$ is a $p \times q$ matrix, the Kronecker matrix product $\bm{A} \otimes \bm{B}$ is
the $mp \times nq$ block matrix
\begin{equation}
\label{eq:KMP}
    \bm{A} \otimes \bm{B} = \left[ {\begin{array}{*{20}c}
                {A_{1,1} \bm{B}} &  \cdots  & {A_{1,n} \bm{B}}  \\
                \vdots  &  \ddots  &  \vdots   \\
                {A_{m,1} \bm{B}} &  \cdots  & {A_{m,n} \bm{B}}  \\
                \end{array} } \right]
\end{equation}

The operator ${\rm vec}(\bm{A})$ denotes the vectorization of the matrix $\bm{A}$ formed by stacking the columns of $\bm{A}$ into a single column vector
\begin{equation}
    {\rm vec}(\bm{A})= \left[ {\begin{array}{*{20}c}
                        {A_{:,1} }  \\
                        \vdots   \\
                        {A_{:,m} }  \\
                        \end{array} } \right]
\end{equation}

For a vector $\bm{x}$ of dimension $n$, the operator ${\rm diag}( \; )$ returns the $n \times n$ diagonal matrix
\begin{equation}
    {\rm diag} (\bm{x})  = \left[ {\begin{array}{*{20}c}
    {x_1 } &  \cdots  & 0  \\
    \vdots  &  \ddots  &  \vdots   \\
    0 &  \cdots  & {x_n }  \\
    \end{array} } \right]
\end{equation}

For a matrix $\bm{A}$ that is $m \times n$, the operator ${\rm diag}(\;)$ returns the diagonal $mn \times mn$ matrix
\begin{equation}
    {\rm diag}({\bm{A}}) = \left[ {\begin{array}{*{20}c}
    {{\rm diag}(A_{:,1} )} &  \cdots  & {{\bm{0}}_n }  \\
    \vdots  &  \ddots  &  \vdots   \\
    {{\bm{0}}_n } &  \cdots  & {{\rm diag}(A_{:,m} )}  \\
    \end{array} } \right]
 \end{equation}

The standard Euclidean norm of a vector $\bm{x}$ and Frobenius norm of a matrix $\bm{A}$ are denoted by
$\left\| \bm{x} \right\|_2$ and $\left\| \bm{A} \right\|_F$, respectively, and defined as follows

\begin{equation}
    \left\| \bm{x} \right\|_2  = \left( {\sum\limits_{i = 1}^n {x_i^2 } } \right)^{\tfrac{1}
    {2}}, \quad \left\| \bm{A} \right\|_F  = \left( {\sum\limits_{i = 1}^n {\sum\limits_{j = 1}^m {A_{i,j}^2 } } } \right)^{\tfrac{1}
    {2}}
\end{equation}

Two of the applications considered in this paper involve compressible flows governed by the Navier-Stokes equations. 
In this case, the fluid state vector is denoted by $\bm{w}$, and the fluid pressure, density, and velocity magnitude
are denoted by $p$, $\rho$, and $u$, respectively. The free-stream conditions are designated by the subscript $\infty$.

\section{Reduction of embedded boundary models}
\label{sec:EBM}

In this work, FSI problems are considered as a vehicle to explain the snapshot compression
method proposed for constructing a Reduced-Order Basis (ROB) suitable for the nonlinear reduction of embedded boundary
models. It is stressed however that this method is general.  In principle, it applies to any EBM framework designed for
solving any computational problem. Before presenting this method in Section~\ref{sec:MET} however, the stage is set
here by describing the generic problem of interest, and its reduction process.

\subsection{Generic nonlinear fluid-structure interaction problem}
\label{sec:FSI_problem}

\begin{figure}
\centering
\includegraphics{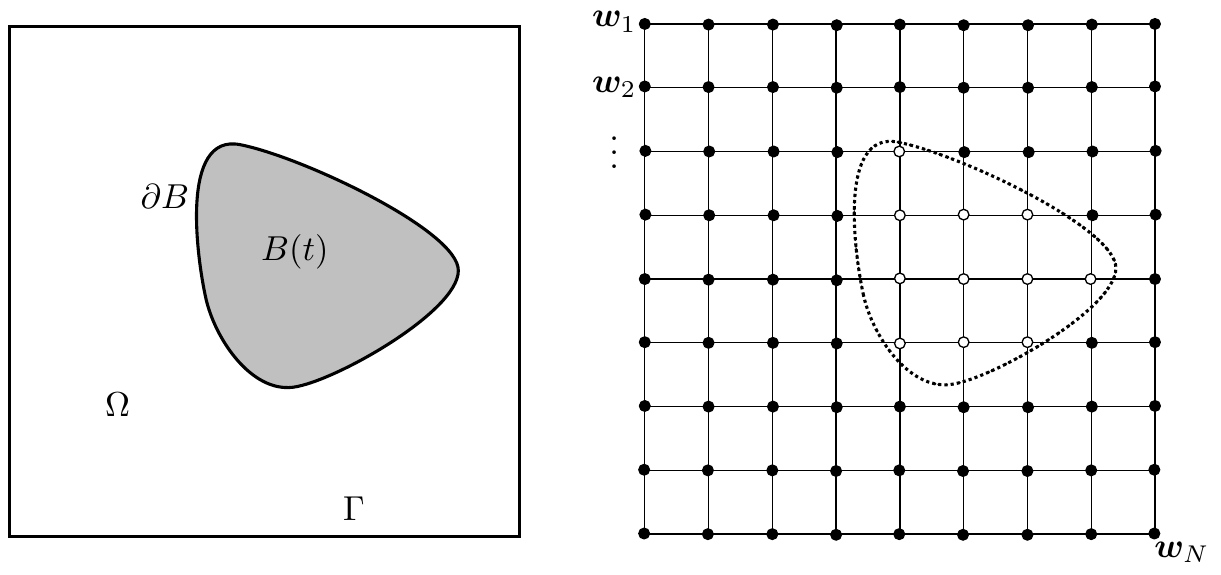}
\caption{Generic setting of an EBM for a generic FSI problem and discretization
(the circles filled in black are referred to as ``real'' nodes, whereas the empty ones are referred to as ``ghost'' nodes}\label{fig:FSI_problem}
\end{figure}

Consider the problem of computing an unsteady compressible flow around a {\it rigidly moving} body $B(t)$ contained in a
{\it fixed} region $\Omega \subset \mathbb{R}^d(d=2,3)$. A representative geometry of this problem is illustrated in the left part
of Figure~\ref{fig:FSI_problem}. Assume that the boundary $\Gamma$ and fluid/body interface $\partial B$ are
equipped with the appropriate boundary and fluid-structure transmission conditions, respectively.
In an EBM framework, a background Eulerian
mesh is typically used to discretize the global domain $\Omega$. The nodes of this mesh that are occluded by $B(t)$ at time $t$
are usually referred to as ``ghost'' nodes; the others are referred to as ``real'' nodes.
A semi-discretization on this mesh of the Navier-Stokes equations governing the fluid subsystem yields
a set of nonlinear ordinary differential equations (ODEs) which can be written as
\begin{equation}
\label{Eqn:semi_dis}
\frac{d\bm{w}}{{dt}} + \bm{f}(\bm{w}) = \bm{0}
\end{equation}
where $t$ denotes time, $\bm{w}(t) \in \mathbb{R}^{cN}$, $c$ is the number of
fluid variables per mesh node, $N$ denotes the total number of mesh nodes, and $\bm{f}:\mathbb{R}^{cN}  \to \mathbb{R}^{cN}$ denotes
a nonlinear function of $\bm{w}$ that represents the convective and diffusive fluxes.

Without any loss of generality, it is assumed throughout the remainder of this paper that Eq.~\eqref{Eqn:semi_dis} is
discretized in time using an implicit linear multistep scheme. Hence, if $t^0 = 0 < t^1 < \cdots < t^{N_t} = T$ denotes a
discretization of the time-interval $[0,T]$ and $\bm{w}^n \approx \bm{w}(t^n), n \in \{1,\cdots,N_t \}$, the discrete
counterpart of Eq.~\eqref{Eqn:semi_dis} at time-step $n \in \{1,\cdots,N_t \}$ is
\begin{equation}
\label{Eqn:full_dis}
\bm{R}(\bm{w}^n ) = \sum\limits_{j = 0}^s {\alpha _j \bm{w}^{n - j} }  + \sum\limits_{j = 0}^s {\beta _j \bm{f}(\bm{w}^{n - j} )}=0
\end{equation}
where $s$ is the order of accuracy of the chosen time-integrator and $\alpha_j$ and $\beta_j$ are two constants
that characterize it.

In an EBM framework for CFD, the fluid-structure transmission conditions require special attention because the fluid/body
interface $\partial B$ does not coincide in general with the nodes of the background mesh. Over the years, a
large number of different boundary treatment schemes has been developed for addressing this issue (for example, see
~\cite{Mittal_ARFM_2005} and~\cite{Zeng12} for a review of this and related issues). This is noteworthy because
the method proposed in this paper for compressing solution snapshots computed by EBMs based on these schemes is
actually independent of them.

\subsection{Nonlinear model reduction}
\label{sec:MOR}
In a projection-based MOR method, the state vector $\bm{w}^n \in \mathbb{R}^{cN}$ is approximated in a global affine trial subspace
as follows
\begin{equation}
    \label{Eqn:tilde_w}
    \bm{w}^n  \approx  \widetilde{\bm{w}} : = \bm{w}^0 + \bm{U} \bm{a}^n
\end{equation}
where $\bm{U}  \in \mathbb{R}^{cN \times k} $ is a matrix whose columns contain the basis of this subspace, $k << cN$ is
the reduced dimension,
$\bm{a}^n \in \mathbb{R}^{k}$ denotes the generalized coordinates in this basis at time $t^n$, and $\bm{w}^0$ is the initial
condition. Substituting Eq.~\eqref{Eqn:tilde_w} into Eq.~\eqref{Eqn:full_dis} yields $\bm{R}(\bm{w}^0 + \bm{U} \bm{a}^n)=0$,
which represents an overdetermined system of $cN$ equations with $k$ unknowns. Consequently, the residual is projected onto a
test subspace represented by $\bm{\Phi} \in \mathbb{R}^{cN \times k}$, yielding the square system
$\bm{\Phi}^{\rm T}\bm{R}(\bm{w}^0 + \bm{U} \bm{a}^n)=0$. In a Galerkin projection, $\bm{\Phi} =\bm{U}$.
In a least-square projection, $\bm{\Phi} = \bm{J} \bm{U}$~\cite{Carlberg1}, where
$J_{:,i}={{\partial R(\bm{w}^0 + \bm{U} \bm{a}^n)} \mathord{\left/ {\vphantom {{\partial R(\bm{w}^0 + \bm{U} \bm{a}^n)} {\partial a_i^n }}} \right. \kern-\nulldelimiterspace} {\partial a_i^n }}$
is the Jacobian of $\bm{R}(\bm{w}^0 + \bm{U} \bm{a}^n)$~(\ref{Eqn:full_dis}) with respect to $\bm{a}$.  In either case,
the vector of generalized coordinates is then obtained by solving a square system of nonlinear equations using Newton's method
or a preferred variant.

Computing the vector of generalized coordinates $\bm{a}^n$ using a least-squares projection approach is equivalent to solving at
each time-step the minimization problem~\cite{Carlberg1}
\begin{equation}
\label{eqn:min_prob}
\underset{\bm{a}^n}{\text{min}} \left\| \bm{R}( \bm{w}^0 + \bm{U} \bm{a}^n ) \right\|_2
\end{equation}
For nonlinear, non self-adjoint problems such as those represented in this case by the set of ODEs~(\ref{Eqn:semi_dis}),
this approach is more robust than its Galerkin counterpart. Hence, its application in the context of
an EBM framework is given here further attention.

Since only a portion of the Eulerian computational domain corresponds to the real flow region, minimizing the norm
of the {\it global} residual vector $\bm{R}(\bm{w}^n)$ --- that is, including its ghost component ---
is neither necessary nor convenient. Instead, let $\bm{m}^n \in \{0,1\}^{cN}$ denote a binary vector identifying the fluid/body ---
or more generally, the ghost/real --- partition of the computational domain at time $t^n$
\begin{equation}
\label{eq:m}
    m_{i}^n  = \left\{ {\begin{array}{*{20}l}
		    {1,} & \hbox{if} ~ i \in \Omega \backslash \overline {B}(t^n)  \\
                    {0,} & \hbox{if} ~ i \in {B}(t^n)  \\
     \end{array} } \right.
\end{equation}
The idea here is that at each time-step, the norm of only the real component of $\bm{R}(\bm{w}^n)$ --- that is, that associated
with the real flow region --- needs to be minimized to obtain the vector of generalized coordinates $\bm{a}^n$ needed to
determine the Reduced-Order Model (ROM) approximation~(\ref{Eqn:tilde_w}). In other words, the idea here is that
in the context of an EBM, the minimization problem~(\ref{eqn:min_prob}) can be reformulated as
\begin{equation}
    \label{eqn:min_prob2}
    \underset{\bm{a}^n}{\text{min}} \left\| \bm{m}^n \odot \bm{R}( \bm{w}^0 + \bm{U} \bm{a}^n ) \right\|_2
\end{equation}

Equation~\eqref{eqn:min_prob2} above is a nonlinear equation. Therefore, it must be solved iteratively using, for example,
the Gauss-Newton method, which can be summarized as

{\it for $i = 1, \ldots , p$  solve}
\begin{equation}
    \label{eqn:full Gauss-Newton}
    \underset{\bm{\Delta a}^{(i)}}{\text{min}} \| \bm{m}^n \odot ( \bm{J}^{(i)} \bm{U} \bm{\Delta a}^{(i)}  + \bm{R}( \bm{U} \bm{a}^{n,(i)} ) ) \|_2
\end{equation}
where the superscript $(i)$ designates the $i$-th iteration, $\bm{\Delta a}^{(i)}$ is the increment of the sought-after solution
at the $i$-th Gauss-Newton iteration, and $p$ is determined by a convergence criterion.

Equation~\eqref{eqn:full Gauss-Newton} constitutes a $k$-dimensional Gauss-Newton ROM of the High-Dimensional Model (HDM) ---
that is, the high-fidelity CFD model --- represented here by Eq.~(\ref{Eqn:semi_dis}). The normal equations associated
with Eq.~\eqref{eqn:full Gauss-Newton} can be written as
\begin{equation}
    \label{eqn:normal}
    (\bm{J}^{(i)} \bm{U})^{\rm T} {\rm diag}(\bm{m}^n ) \bm{J}^{(i)} \bm{U} \bm{\Delta a}^{(i)}  =  - (\bm{J}^{(i)} \bm{U})^{\rm T} {\rm diag}(\bm{m}^n ) \bm{R}( \bm{U} \bm{a}^{n,(i)} )
\end{equation}
where the superscript ${\rm T}$ denotes the transpose.

In any case, whether a Galerkin or least-squares approach is chosen for determining the vector of generalized coordinates
$\bm{a}^n$, the time-invariant ROB $\bm{U}$ must be first determined. In MOR applications,
basis functions are usually constructed from solution snapshots. These snapshots are collected, assembled into a matrix,
and compressed using, for example, SVD. In an interface-fitted computational framework, this computation is straightforward
because the computational domain does not contain a ghost region. In an EBM framework however, this compuation
is complicated by the presence and evolution of a ghost/real partition of the computational fluid domain. In the following
section, a method for constructing optimal fluid basis functions from snapshots computed by an EBM is presented.

\section{Construction of a reduced-order basis for an embedded boundary model}
\label{sec:MET}

To this effect, this section is organized in three parts. First, the concepts of solution snapshots and
a snapshot matrix are recalled. Then, the problem of constructing an optimal ROB for embedded boundary models is formulated
as a weighted low-rank matrix approximation problem. Finally, an efficient iterative scheme for solving this problem is specified.

\subsection{Eulerian snapshots}
\label{sec:Eulerian snapshots}
A solution snapshot, or simply a snapshot, is defined here as a state vector $\bm{w}^n \in \mathbb{R}^{cN}$ computed as the
solution of Eq.~(\ref{Eqn:full_dis}) for some instance of its parameters --- that is, for some specific time $t^n$ or specific
value of the set of flow parameters or boundary/initial conditions underlying this governing equation ---
on the background Eulerian mesh discretizing the domain $\Omega$.
A snapshot matrix is defined as a matrix $\bm{X} \in \mathbb{R}^{cN \times K}$ ($K>k$) whose columns are individual snapshots.
For all practical purposes, the main focus of this paper is on unsteady flows and on snapshots associated with different
time-instances $t^m$. Hence,  $X_{:,i} := \bm{w}^i$ for $i = 1,\cdots,K$.

Because the proposed method for computing a ROB for an embedded boundary model does not utilize information from the
occluded region of the computational domain, the ghost components of these snapshots can take arbitrary values.
Consequently, the proposed method, which is described below, is independent of the specifics of the EBM framework used
to generate the snapshots.

\subsection{Weighted low-rank matrix approximation problem}
\label{sec:WLRA}

Constructing basis functions from snapshots in the spirit of the POD method --- that is, using data compression ---
can be formulated mathematically as a low-rank matrix approximation problem as follows.

{\it For a given snapshot matrix} $\bm{X}~\in~\mathbb{R}^{cN \times K}${\it , find a lower rank matrix}
$\widetilde{\bm{X}}~\in~\mathbb{R}^{cN \times K}$ {\it that solves the minimization problem}
\begin{equation}
 \label{Eqn:LRA}
    \underset{{\rm rank}(\widetilde{\bm{X}})=k}{\text{min}}
 \displaystyle \| \bm{X} - \widetilde{\bm{X}} \|_F
\end{equation}
where $k < K \ll cN$. In this problem, the rank constraint can be taken care of by representing the unknown matrix
as $\widetilde{\bm{X}}=\bm{U} \bm{V}$, where $\bm{U}~\in~\mathbb{R}^{cN \times k}$ and $\bm{V}~\in~\mathbb{R}^{k \times K}$,
so that problem~(\ref{Eqn:LRA}) becomes
\begin{equation}
\label{Eqn:LRA2}
\underset{\bm{U} \in \mathbb{R}^{cN \times k},\bm{V} \in \mathbb{R}^{k \times K}}{\text{min}}
\displaystyle \| \bm{X} - \bm{U}\bm{V} \|_F
\end{equation}
It is well-known that the solution of the above low-rank approximation problem is given by the SVD of $\bm{X}$: specifically,
$\bm{U}=U_{:,1:k}^*$ and $\bm{V}=(\bm{\Sigma} \bm{V}^{*{\rm T}})_{1:k,:}$ where
$\bm{X}=\bm{U}^* \bm{\Sigma} \bm{V}^{*{\rm T}}$.

Unfortunately, when the snapshot matrix $\bm{X}$ is generated using an EBM framework, the solution outlined above
cannot be expected to yield an optimal ROB. This is because: (a) the snapshots computed using an EBM contain information
from both the flow (or real) region of the computational domain and its occluded (or ghost) region, and (b) this data
inconsistency is not accounted for in the standard low-rank matrix approximation problem~(\ref{Eqn:LRA}).
Hence, this issue is resolved here by proposing the alternative weighted low-rank matrix approximation problem
\begin{equation}
 \label{Eqn:WLRA}
 \underset{\bm{U} \in \mathbb{R}^{cN \times k},\bm{V} \in \mathbb{R}^{k \times K}}{\text{min}}
 \displaystyle \| \bm{M} \odot (\bm{X} - \bm{U}\bm{V}) \|_F
\end{equation}
where $\bm{M}\in \{0,1\}^{cN \times K}$ is a binary mask matrix identifying the ghost/real partition of the snapshot matrix
\begin{equation}
    \bm{M}: = \left[ {\begin{array}{*{20}c} {\bm{m}^0 } & {\bm{m}^1 } &  \cdots  & {\bm{m}^K }  \end{array} } \right]
\end{equation}
where $\bm{m}^l$ has already been defined in~(\ref{eq:m}).
This binary weighting ensures that the values of $\bm{w}$ at the ghost nodes do not play a role in the construction of
the basis functions,
which in turn implies that the derived basis functions are optimal for the real components of the snapshots
associated with $\Omega \backslash \overline {B}$.

In the general case (i.e. ${\rm rank}(\bm{M})>1$), the weighted low-rank approximation problem formulated above is not reducible
to the un-weighted problem. Therefore, its solution is not given by the SVD factorization of the snapshots~\footnote{The weighted low-rank matrix approximation problem is reducible to the un-weighted problem only for the special case where ${\rm rank}(\bm{M})=1$;
see~\ref{appB} for details.}. Furthermore, no closed form solution of this problem is known. Hence, it must be solved by numerical
iterations. In this work, the Alternating Least Squares (ALS) algorithm is chosen for this purpose. This algorithm is the most
widely applicable and empirically successful approach for solving this and related problems~\citep{Chen_2008,Okatani_2007,Mitra_2010}.

\subsection{Alternating least squares algorithm}
\label{sec:ALS}

The ALS algorithm takes advantage of the bi-linearity of the representation $\widetilde{\bm{X}}=\bm{U} \bm{V}$.
Using the connection between the Frobenius and Eucledian norms and the notation for the Kronecker matrix product~(\ref{eq:KMP}),
problem~\eqref{Eqn:WLRA} can be re-written as
\begin{subequations}
\begin{align}
	&\underset{\bm{U} \in \mathbb{R}^{cN \times k},\bm{V} \in \mathbb{R}^{k \times K}}{\text{min}} \| \bm{M} \odot (\bm{X} - \bm{U}\bm{V}) \|_F \nonumber\\
	&\Leftrightarrow  \underset{\bm{U} \in \mathbb{R}^{cN \times k},\bm{V} \in \mathbb{R}^{k \times K}}{\text{min}} \left\| {{\rm diag}(\bm{M})\left( {{\rm vec}(\bm{X}) - {\rm vec}(\bm{UV})} \right)} \right\|_2 \\
      &\Leftrightarrow  \underset{\bm{U} \in \mathbb{R}^{cN \times k},\bm{V} \in \mathbb{R}^{k \times K}}{\text{min}} \left\| {{\rm diag}(\bm{M})\left( {{\rm vec}(\bm{X}) - (\bm{I}_K \otimes \bm{U}){\rm vec}(\bm{V})} \right)} \right\|_2 \label{Eqn:LLS1} \\
      &\Leftrightarrow  \underset{\bm{U} \in \mathbb{R}^{cN \times k},\bm{V} \in \mathbb{R}^{k \times K}}{\text{min}} \left\| {{\rm diag}(\bm{M})\left( {{\rm vec}(\bm{X}) - (\bm{V}^{\rm T}  \otimes \bm{I}_{cN} ){\rm vec}(\bm{U})} \right)} \right\|_2 \label{Eqn:LLS2}
    \end{align}
\end{subequations}
Problem~\eqref{Eqn:LLS1} is a linear least-squares problem for $\bm{V}$ if $\bm{U}$ is known. Problem~\eqref{Eqn:LLS2} is a linear
least-squares problem for $\bm{V}$ if $\bm{U}$ is known. This suggests the following simple iterative solution procedure:
\begin{enumerate}
	\item Guess $\bm{U} \in \mathbb{R}^{cN \times k}$ (for example, initialize $\bm{U}$ as the first $k$ left-singular vectors
		of $\bm{X}$)
  \item Repeat until convergence
      \begin{enumerate}
	      \item Solve problem~\eqref{Eqn:LLS1} for $\bm{V} \in \mathbb{R}^{k \times K}$ given $\bm{U} \in \mathbb{R}^{cN \times k}$
	      \item Solve problem~\eqref{Eqn:LLS2} for $\bm{U} \in \mathbb{R}^{cN \times k}$ given $\bm{V} \in \mathbb{R}^{k \times K}$
        \end{enumerate}
\end{enumerate}

A more detailed outline of the above algorithm is given in Algorithm~\ref{Alg:AP}, where the stopping criterion has
been omitted for the sake of brevity (see~\citep{Markovsky_book_2012,Markovsky_JCAM_2013} for a discussion of this topic).

\begin{algorithm}
    \caption{Alternating least squares algorithm for the solution of the weighted low-rank approximation problem \label{Alg:AP}}
    \begin{algorithmic}[1]
        \REQUIRE Data matrix $\bm{X} \in \mathbb{R}^{cN \times K}$ and weighting matrix $\bm{M} \in \mathbb{R}^{cN \times K}$
        \ENSURE Locally optimal solution $\left[ \bm{U}^{(i)}, \bm{V}^{(i)}\right]$ of problem~\eqref{Eqn:WLRA}
        \STATE Populate the ghost entries of the snapshot matrix $\bm{X}$ with zeros
        \STATE Compute the SVD of the snapshot matrix: $\bm{X}=\bm{U}^* \bm{\Sigma} \bm{V}^{*{\rm T}}$
          \item Initialize: $\bm{U}^{(0)} = U_{:,1:k}^*$
        \FOR{$i=1,2,\ldots$ until convergence}
            \STATE Solve problem~\eqref{Eqn:LLS1}, i.e. evaluate
                    \begin{equation*}
                        \label{eqn:V_U}
                        {\rm vec}(\bm{V}^{(i)} ) = ({\rm diag}(\bm{M})(\bm{I}_{K} \otimes \bm{U}^{(i)}))^{+}{\rm diag}(\bm{M}){\rm vec}(\bm{X})
                     \end{equation*}
            \STATE Solve problem~\eqref{Eqn:LLS2}, i.e. evaluate
                    \begin{equation*}
                        \label{eqn:U_V}
                        {\rm vec}(\bm{U}^{(i+1)} ) = ({\rm diag}(\bm{M})(\bm{V}^{{(i)} \rm T} \otimes \bm{I}_{cN}))^{+}{\rm diag}(\bm{M}){\rm vec}(\bm{X})
                    \end{equation*}
        \ENDFOR
    \end{algorithmic}
\end{algorithm}

The superscripts of $\bm{U}$ and $\bm{V}$ in Algorithm~\ref{Alg:AP} designate an ALS iteration.
The computational cost of this algorithm is roughly $k$ times the computational cost of a single thin SVD
of the snapshot matrix times the number of iterations for convergence --- that is, ${\rm O}\left( {cNk^2p} \right)$,
where $p$ denotes the performed number of iterations. In Section~\ref{sec:APP}, it is shown that in general,
$p = 1$ suffices to obtain good results. The convergence of this algorithm is monitored using the normalized
weighted projection error
\begin{equation}
\label{eqn:projection_error}
    e^{(i)} := \| \bm{M} \odot (\bm{X} - \bm{U}^{(i)} \bm{V}^{(i)})  \|_F^2 / \| \bm{M} \odot \bm{X}  \|_F^2
\end{equation}
This error is guaranteed to decrease monotonically to a local minimum~\citep{Markovsky_book_2012,Markovsky_JCAM_2013}.

A MATLAB implementation of Algorithm~\ref{Alg:AP} is provided in~\ref{appA}.

\section{Applications}
\label{sec:APP}

Three FSI problems are considered here to illustrate the method proposed in this paper for constructing
basis functions suitable for the nonlinear model reduction of embedded boundary models, and demonstrate its features
and performance. The first one is a simple one-dimensional model problem based on Burgers' equation. It has the merit of
illustrating the basic idea and being easily reproducible by the interested reader. The second problem focuses on the prediction
of a two-dimensional unsteady viscous flow around a cylindrical body in large-amplitude heaving motion.
Because this problem is formulated in an
unbounded fluid domain, it is also solvable using an interface- (or body-) fitted
ALE framework. Hence, it offers a venue not only for assessing the intrinsic
performance of the proposed method for constructing a suitable ROB for an embedded boundary model that is more representative
than that of the previous example, but also assessing the performance of the resulting nonlinear ROM relative to that of a
counterpart ROM constructed using a body-fitted computational framework.
The third problem focuses on the solution of a two-dimensional unsteady turbulent flow inside a
square cavity containing a rotating ellipsoidal body. In this case, the large-amplitude motion of $B(t)$ challenges the
robustness if not feasibility of a body-fitted ALE framework and calls instead for an EBM framework. Hence, this third problem
highlights the need for and demonstrates the performance of the computational methodology proposed in this paper.

For each FSI problem outlined above, the governing (fluid) equations are semi-discretized using the central
finite difference method. They are discretized in time using the backward Euler implicit scheme with a constant
time-step $\Delta t$ chosen so that $B(t)$ does not travel more than one layer of nodes during this time-step.
The Newton method is used to solve all nonlinear equations arising from the implicit time-discretization.

Whenever the flow problem of interest is solved using a body-fitted ALE framework, a ROB for this problem
is constructed by computing $K$ snapshots of the solution at different time instances $t^n$,
then compressing them using the SVD. In this case, the nonlinear ROM of interest is generated using the GNAT
method~\cite{Carlberg1,Carlberg_JCP_2013} based on a least-squares projection.

On the other hand, whenever an EBM framework is used to solve the flow problem of interest, a ROB for this problem
is constructed by computing $K$ snapshots of the solution at different time instances $t^n$,
and compressing them using the ALS algorithm described in Algorithm~\ref{Alg:AP}. In this case, the nonlinear ROM of interest
is constructed using a variant of the same GNAT method where at each time-step, the vector of generalized coordinates is
computed by solving~(\ref{eqn:min_prob2}).

\subsection{One-dimensional fluid-structure model problem based on the viscous Burgers equation}
\label{sec:Burgers}

First, the following one-dimensional FSI problem based on the viscous Burgers equation, a periodic boundary condition
for the fluid subsystem, and non-homogeneous Dirichlet boundary conditions for the body subsystem in lieu of the fluid-structure
transmission conditions is considered
\begin{equation}
    \label{eqn:Burgers_eq}
\left \{
	\begin{array}{l c l}
    \displaystyle{\frac{{\partial w}}
    {{\partial t}} + w\frac{{\partial w}}
    {{\partial x}}} &=& \displaystyle{\frac{1}
{{\operatorname{Re} }} \frac{{\partial ^2 w}} {{\partial x^2 }}, \quad (x,t) \in [0,1] \times [0,2]}\\
    w(x,0) &=& 0\\
    w\left( {0,t} \right) &=& w\left( {1,t} \right)\\
 w\left( {\xi(t),t} \right) &=& w\left( {\xi(t) + 0.1,t} \right) = \sin (2 \pi t)
\end{array}
\right .
\end{equation}
where $\operatorname{Re}$ is a Reynolds-like number and is set to $\operatorname{Re} = 500$, and
\begin{equation}
\xi(t) = (1 - 0.1)/2 + 0.1 \sin (2 \pi t)
\end{equation}

The above initial boundary value problem models a flow problem in an unbounded, one-dimensional, fluid domain in the middle of
which a rigid, linear body $B$ of length $0.1$ is placed and set into the oscillatory harmonic motion characterized by
the amplitude 0.1, frequency $2\pi$, and velocity $\displaystyle{d\xi/dt} = 0.2\pi \cos (2\pi t)$.
At each time $t$, $x = \xi(t)$ defines the position of the left
extremity of $B(t)$, and $x = \xi(t) + 0.1$ defines the position of its right extremity.
Hence, this problem is a one-dimensional instance of the generic FSI problem discussed in Section~\ref{sec:FSI_problem}.

The simplicity of problem~(\ref{eqn:Burgers_eq}) is such that it can be solved using both an ALE framework
on a moving mesh, and an EBM framework on a fixed mesh. Hence, both frameworks are considered here, primarily for the
purpose of comparisons. In both cases, Eq.~\eqref{eqn:Burgers_eq} is discretized on a uniform Cartesian mesh
with $\Delta x = 0.001$ ($N = 1,000$ nodes).

\begin{figure}
    \centering
    \includegraphics{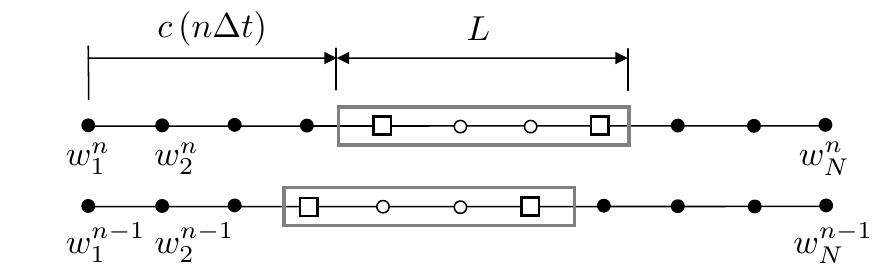}
    \caption{EBM framework for the solution of the one-dimensional FSI problem based on the viscous Burgers equation}
    \label{fig:Burgers_EBM}
\end{figure}

\begin{figure}
\centering
    \includegraphics{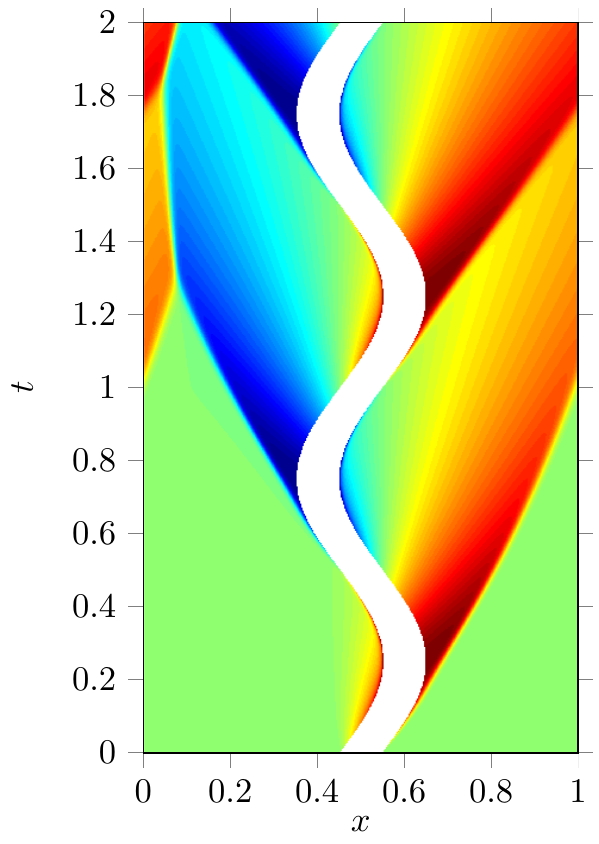}
\caption{Solution of the one-dimensional FSI problem based on the viscous Burgers equation}\label{fig:Burgers}
\end{figure}

Specifically, due to the unbounded fluid domain assumption and the periodic fluid boundary condition, the body-fitted ALE
framework considered here for the solution of problem~(\ref{eqn:Burgers_eq}) is equipped with a rigid motion of the mesh
that is identical to that of $B(t)$. As for the considered EBM framework, it is equipped with a first-order ghost fluid
scheme where the ghost values of $w$ are populated at each time-instance $t^n = n \Delta t$ only
at the two ghost fluid nodes that are the nearest to the left and right boundaries of $B(t^n)$, using the same
value $w^n =\sin (2 \pi n \Delta t)$. This EBM framework is illustrated in Figure~\ref{fig:Burgers_EBM}, where the
circles filled in black represent the real nodes, the empty ones represent the ghost fluid nodes, and the empty squares
designate the subset of ghost fluid nodes whose ghost fluid values are populated. Both computational frameworks deliver
the same HDM solution which is graphically depicted in Figure~\ref{fig:Burgers}.

For each of the body-fitted ALE and EBM models, three ROBs of dimension $k = 10$, 20, and 40 are constructed using $K = 2,000$
snapshots of the solution of problem~(\ref{eqn:Burgers_eq}), and three corresponding nonlinear ROMs are generated.

\begin{figure}
\centering
\subfigure[Variation of the projection error $e^{(i)}$~\eqref{eqn:projection_error} with the $i$-th ALS iteration\label{sufig:projectin_error_burgers}]
    {
        \includegraphics[scale=1]{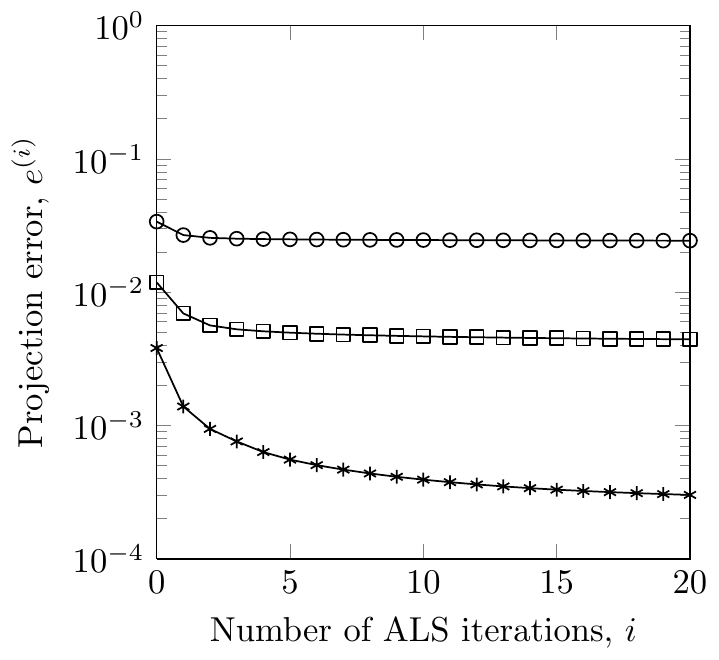}
    }
    \subfigure[Variation of the EBM ROM error $e_{ROM}^{(i)}$ with the $i$-th ALS iteration: the grey horizontal lines indicate
    the errors of the approximate solutions based on the body-fitted ALE ROMs of same dimensions\label{sufig:ROM_error_burgers}]
    {
        \includegraphics[scale=1]{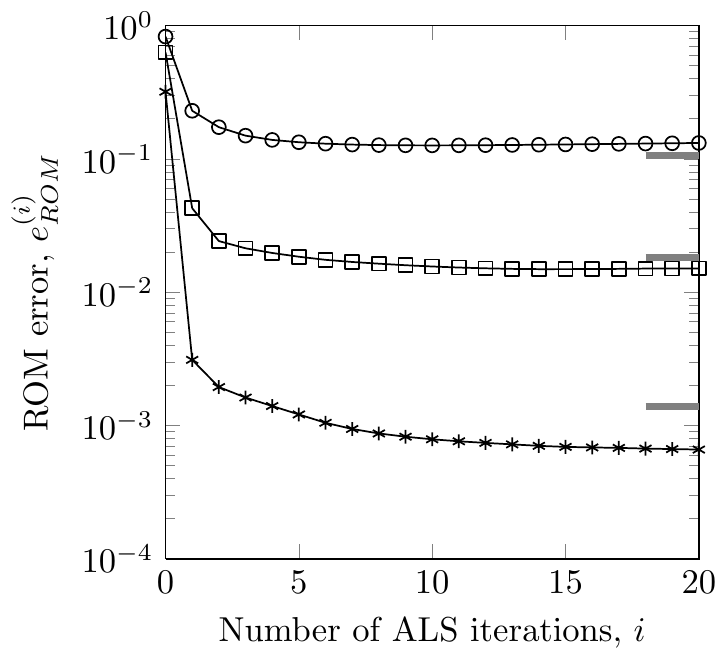}
    }
    \caption{Convergence of the ALS algorithm and associated EBM ROM solutions for the model FSI problem based on the viscous Burgers equation: $k=10$ (circles), $k=20$ (squares), $k=40$ (asterisks)}\label{fig:Burgers_convergence}
\end{figure}

Figure~\ref{sufig:projectin_error_burgers} reports on the convergence of the ALS algorithm using the normalized
weighted projection error~\eqref{eqn:projection_error}. It illustrates a well-known property of this algorithm,
namely, its guarantee to deliver a monotonic convergence to a local minimum.
Figure~\ref{sufig:ROM_error_burgers} reports on the variation with the ROM dimension $k$ of the EBM ROM error
$e_{ROM}^{(i)}$ defined as
\begin{equation}
    \label{eqn:ROM_error_def}
    e_{ROM}^{(i)} := \| \bm{M} \odot (\bm{X} - \bm{U}^{(i)}\bm{A})  \|_F^2 / \| \bm{M} \odot \bm{X}  \|_F^2
\end{equation}
where $\bm{X} \in \mathbb{R}^{N \times K}$ is the matrix of snapshot solutions of problem~(\ref{eqn:Burgers_eq}),
$\bm{U}^{(i)} \in \mathbb{R}^{N \times k}$ is the ROB computed at the $i$-the iteration of the ALS algorithm, and
$\bm{A} \in \mathbb{R}^{k \times K}$ is the matrix of generalized coordinates of the approximate solutions associated
with $\bm{U}^{(i)}$ and can be written as
\begin{equation}
	    \bm{A} : = \left[ {\begin{array}{*{20}c} \bm{a}^0  & \bm{a}^1  &  \cdots  & \bm{a}^K  \end{array} } \right]
\end{equation}
--- that is, each $j$-th column of the matrix product $\bm{U}^{(i)}\bm{A}$ is the EBM ROM solution
of problem~(\ref{eqn:Burgers_eq}) at time $t^j = j \Delta t$, based on the ROB $\bm{U}^{(i)}$ computed at the $i$-th iteration
of Algorithm~\ref{Alg:AP}.

\begin{figure}
    \centering
        \subfigure[$k = 10$: body-fitted ROM (left), EBM ROM (right)\label{subfig:Burgers_n10}]
            {
            \includegraphics{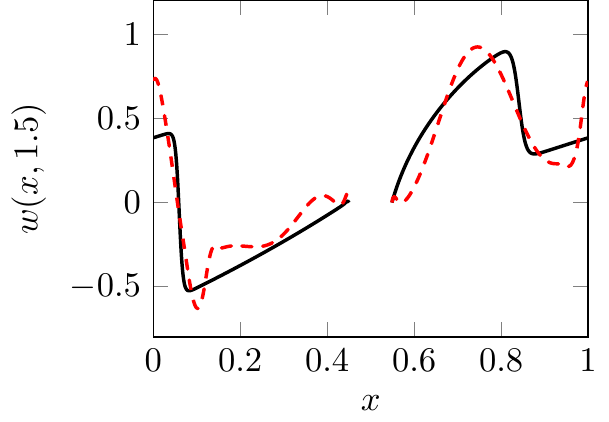}
            \includegraphics{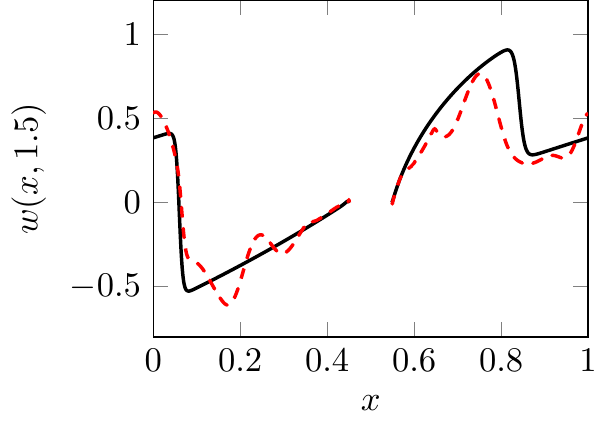}
            }
        \subfigure[$k = 20$: body-fitted ROM (left), EBM ROM (right)\label{subfig:Burgers_n20}]
            {
            \includegraphics{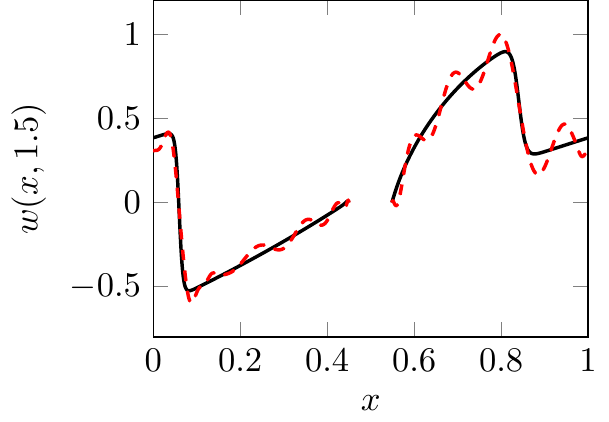}
            \includegraphics{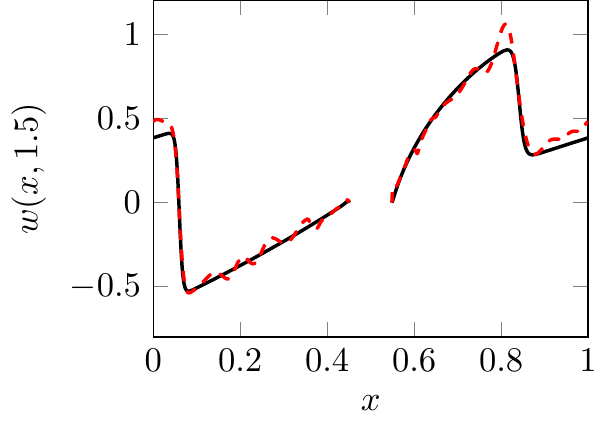}
            }
        \subfigure[$k = 40$: body-fitted ROM (left), EBM ROM (right)\label{subfig:Burgers_n40}]
            {
            \includegraphics{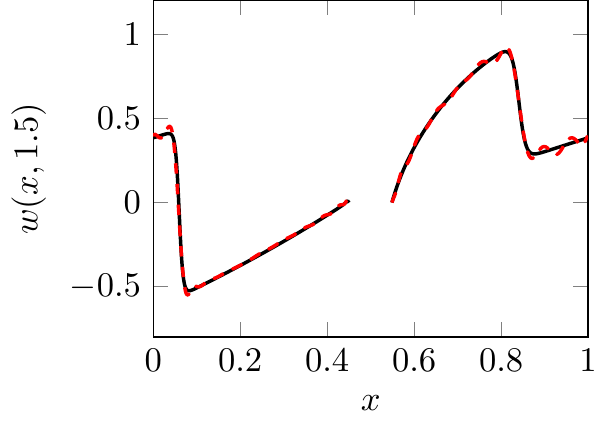}
            \includegraphics{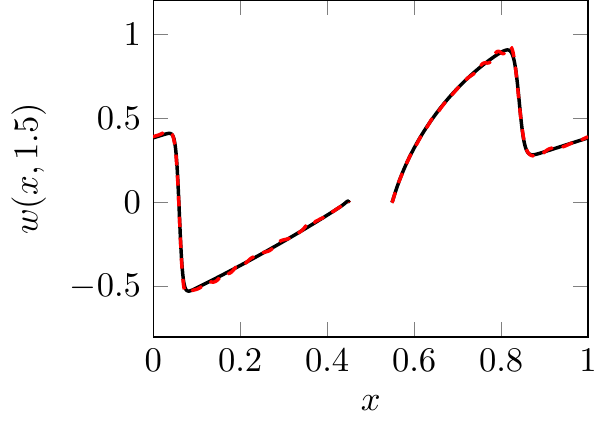}
            }
    \caption{EBM ROM vs. ALE ROM solutions at $t=1.5$ for the model FSI problem based on the viscous Burgers equation:
    reference HDM solution (solid), ROM solution (dashed)}\label{fig:Burgers2}
\end{figure}

At the initial ($0$-th) ALS iteration, the ROB $\bm{U}^{(0)}$ is constructed by performing the SVD directly on EBM solution
snapshots where the ghost values of the solution --- that is, the values of this solution at the occluded mesh nodes ---
are populated with zeros. As shown in Figure~\ref{sufig:ROM_error_burgers}, this approach --- referred to here as
the ``naive approach'' --- delivers a poor performance, which illustrates the need for an alternative method
for constructing EBM ROBs. Figure~\ref{sufig:ROM_error_burgers} also reveals that a single ALS iteration
significantly improves the computed EBM ROM solution.
The results reported in Figure~\ref{sufig:ROM_error_burgers} also show that after a single ALS iteration,
the constructed EBM ROMs deliver already a comparable accuracy to that of the body-fitted ALE ROMs of same dimensions.

Figure~\ref{fig:Burgers2} displays the convergence of the EBM ROM solutions
at $t=1.5$ for $\bm{U} = \bm{U}^{(1)}$, and contrasts these solutions with their counterparts obtained using the body-fitted
ALE ROMs of same dimensions. The reader can observe that both types of ROMs exhibit comparable convergence behavior and accuracy,
thereby demonstrating the effectiveness of the proposed method for constructing suitable EBM ROBs.

\subsection{Two-dimensional fluid-structure interaction problem in an unbounded fluid domain}
\label{sec:cylinder}

\begin{figure}
\centering
\includegraphics{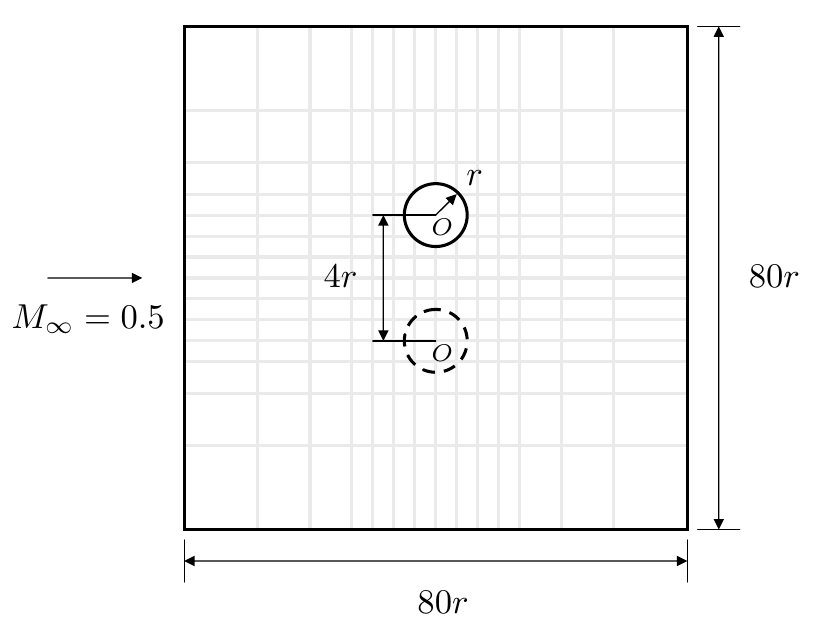}
\caption{FSI problem with a heaving cylindrical body}\label{fig:Cylinder_geometry}
\end{figure}

Next, the simulation of a compressible viscous flow around a heaving rigid cylindrical body $B(t)$
with a circular cross section of radius $r = 0.5$ is considered. The fluid is assumed to be a perfect gas
with the ratio of specific heats $\gamma = 1.4$ and is modeled using the two-dimensional
compressible Navier-Stokes equations. The cylinder is assumed to be infinitely long,
and therefore the flow is effectively modeled as a two-dimensional problem around a disk. The physical fluid domain
is assumed to be unbounded, but the computational fluid domain is bounded by a square of edge length equal to
$80r = 40$ as shown in Figure~\ref{fig:Cylinder_geometry}. The free-stream Mach number is set to $M_{\infty} = 0.5$, 
the Reynolds number based on the cylinder diameter to $\operatorname{Re} = 200$, and the Prandtl number to ${\rm Pr} = 0.72$.

The center of the disk representing $B(t)$, $O$, is set into the vertical sinusoidal motion
\begin{equation}
\label{eq:RM}
y_O(t) = 2 r \sin(t) =  \sin(t)
\end{equation}
where $y_O$ denotes the $y$ abscissae of $O$.
Hence, the total center-to-center displacement of $B(t)$ is $4 r = 2$, which is a relatively large displacement.
Yet, because the fluid domain is also unbounded in this case, and because $B$ is rigid, the FSI problem outlined
above can be reliably solved by a body-fitted ALE framework where the CFD mesh is rigidly moved to follow
the vertical oscillations of $B(t)$.

Hence, two similar but strictly speaking different meshes are constructed for discretizing the spatial domain shown 
in Figure~\ref{fig:Cylinder_geometry}. The first mesh is designed for the EBM framework. It is non-uniform, has
$256 \times 256 = 65,536$ elements whose size in the vicinity of the cylindrical body is uniform and at its finest
is characterized by $\Delta x_{min} = \Delta y_{min} = 0.02 \cdot r$, and embeds the circular body $B(t)$.
The second mesh is similar except for the fact that it is a body-fitted mesh.

\begin{figure}
    \centering
        \includegraphics{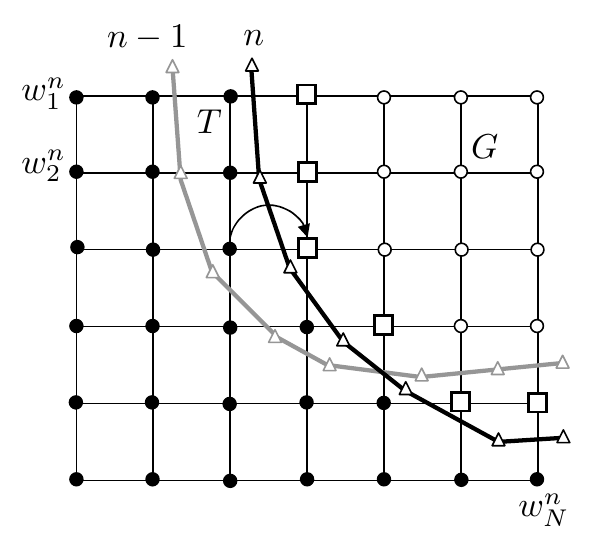}
    \caption{EBM framework for the FSI problem with a heaving cylindrical body: populating the ghost fluid values}\label{fig:Navier-Stokes_EBM} \end{figure}

\begin{figure}
\centering
    \subfigure[Vorticity]
        {
            \includegraphics{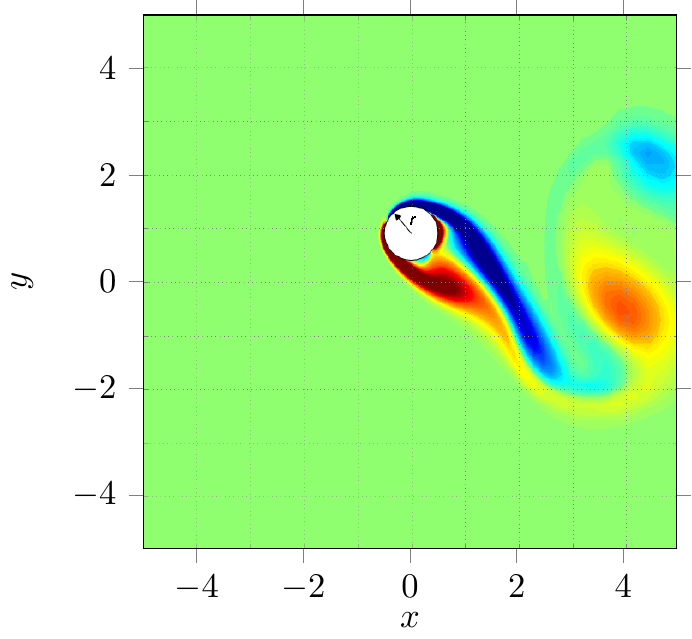}
        }
    \subfigure[Pressure fluctuations]
        {
            \includegraphics{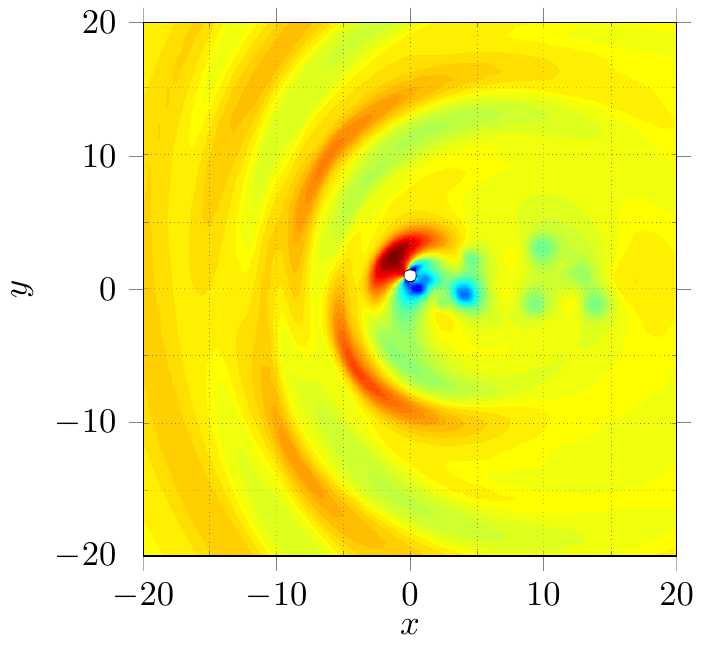}
        }
\caption{Computational snapshots for the FSI problem with a heaving cylindrical body obtained using an EBM}\label{fig:Cylinder}
\end{figure}

The EBM for CFD used for solving this FSI problem is illustrated in Figure~\ref{fig:Navier-Stokes_EBM},
where the real fluid nodes are shown as solid circles and the ghost fluid nodes as empty circles. The kinematic transmission
condition (or adherence condition) on the fluid/body interface is enforced using a first-order ghost-fluid scheme.
At each time-step, the ghost fluid values at the ghost fluid nodes that are the nearest to the fluid/body interface
(these are identified by squares in Figure~\ref{fig:Navier-Stokes_EBM}), are populated as follows. The ghost fluid velocities
at these nodes are set to the velocity of the immersed body $B(t)$. The ghost fluid pressures and densities at these same nodes
are computed by constant extrapolation of their counterparts at the nearest neighboring real fluid nodes.
On the other hand, the ghost fluid values at all other ghost nodes, which are labelled ``G'' in Figure~\ref{fig:Navier-Stokes_EBM},
are not populated.
At the end of each computational time-step, the fluid state vector at the ``transitional'' nodes
(labelled ``T'' in Figure~\ref{fig:Navier-Stokes_EBM}) --- that is, the nodes whose status switches from ghost to real ---
are populated using the ghost fluid values at these nodes from the previous time-step.
Snapshots of the vorticity and pressure fluctuations computed using this EBM and the HDM are shown in Figure~\ref{fig:Cylinder}.

First, a pair of unsteady simulations are performed for $t \in [0,20]$ using the body-fitted ALE and EBM frameworks outlined above,
in order to generate in each case $K = 1,000$ snapshots for model reduction. Then, ROBs and nonlinear ROMs are constructed
as outlined above.

\begin{figure}
\centering
\subfigure[Projection error, $e^{(i)}$ vs. ALS iteration, $i$ for heaving cylindrical body problem\label{sufig:projectin_error_cylinder}]
    {
        \includegraphics{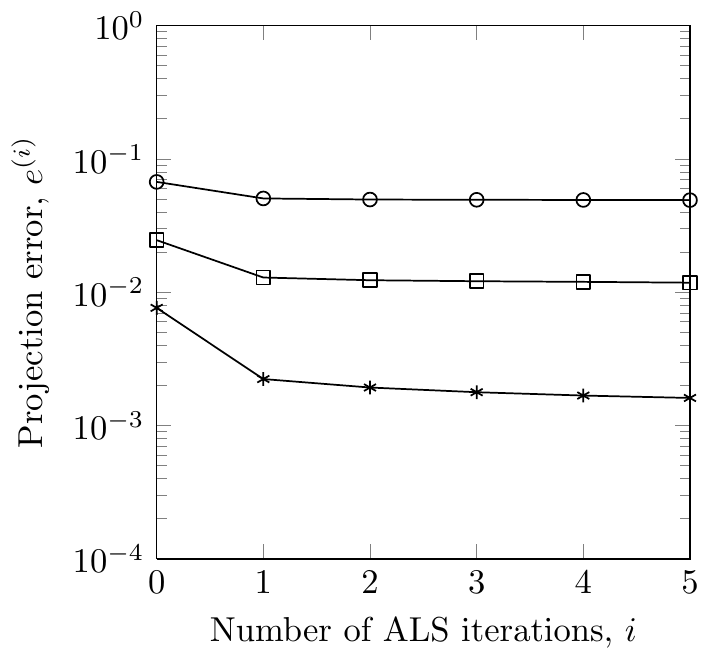}
    }
    \subfigure[Variation of the EBM ROM error $e_{L}^{(i)}$ with the $i$-th ALS iteration: the grey horizontal lines indicate
    the errors of the approximate solutions based on the body-fitted ALE ROMs of same dimensions\label{sufig:ROM_error_cylinder}]
    {
        \includegraphics{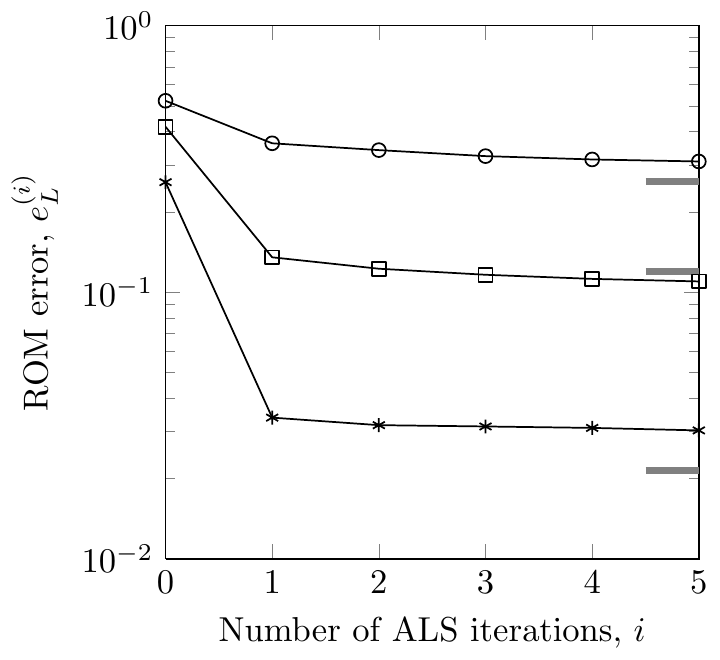}
    }
    \caption{Convergence of the ALS algorithm and associated EBM ROM solutions for the FSI problem with a heaving cylindrical body: $k=10$ (circles), $k=20$ (squares), $k=40$ (asterisks)}\label{fig:Cylinder_convergence}
\end{figure}

Figure~\ref{sufig:projectin_error_cylinder} reports on the convergence of the ALS algorithm
using the projection error $e^{(i)}$~\eqref{eqn:projection_error}. As expected, this error decreases monotonically. 

For this problem, the ROM error is assessed using the relative error in the lift coefficient measured in the $L_2$ norm
\begin{equation}
    e_L ^{(i)}  := \| {C_L ^{CFD}  - C_L ^{ROM} } \|_2 / \| {C_L ^{CFD} } \|_2
\end{equation}
where $C_L : = F_y /\tfrac{1}{2}\rho u_\infty ^2 d$ and $\rho u_\infty ^2 d=1$. This error is reported
in Figure~\ref{sufig:ROM_error_cylinder}. As before, the ROB $\bm{U}^{(0)}$ obtained at the $0$-th iteration of the ALS
algorithm is computed using the so-called naive approach. Figure~\ref{sufig:ROM_error_cylinder} shows that as expected,
the nonlinear ROM constructed using
this ROB performs poorly. However, after only one ALS interation, the ROM constructed using the ROB $\bm{U}^{(1)}$
delivers a good performance. The results reported in Figure~\ref{sufig:ROM_error_cylinder} also demonstrate that
after a single ALS iteration, the generated EBM ROMs deliver an accuracy that is comparable to that of their ALE
ROM counterparts of the same dimensions.

\begin{figure}
    \centering
        \subfigure[$k = 10$; ALE ROM (left) and EBM ROM (right)]
            {
                \includegraphics{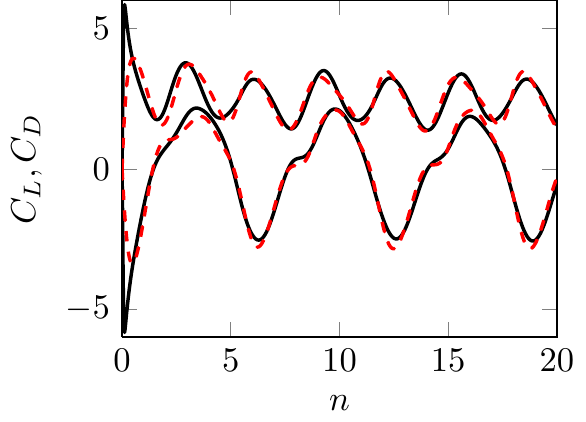}
                \includegraphics{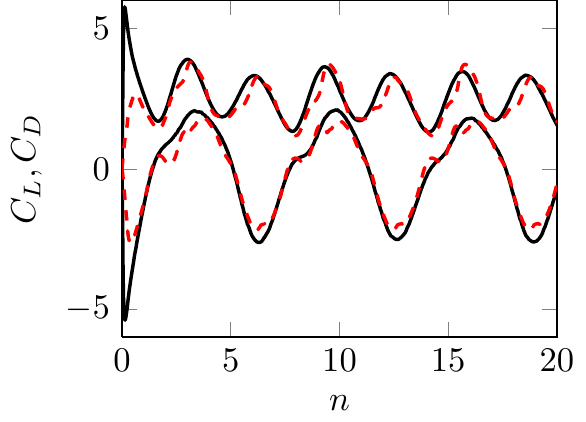}
            }
        \subfigure[$k = 20$; ALE ROM (left) and EBM ROM (right)]
            {
                \includegraphics{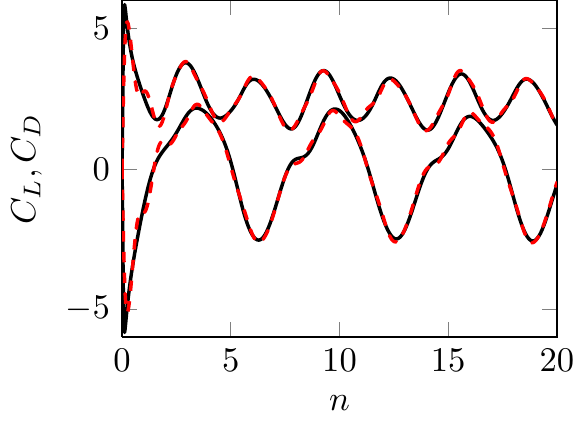}
                \includegraphics{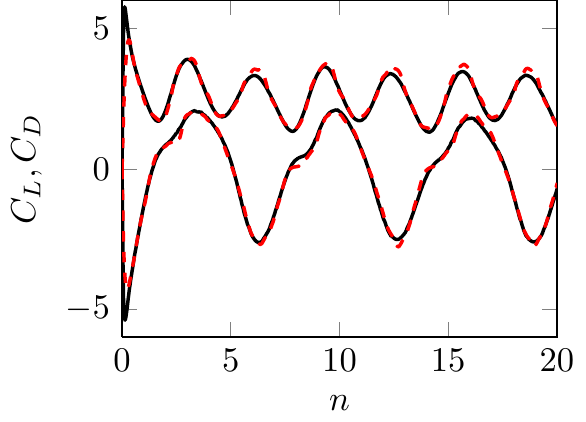}
            }
        \subfigure[$k = 40$; ALE ROM (left) and EBM ROM (right)]
            {
                \includegraphics{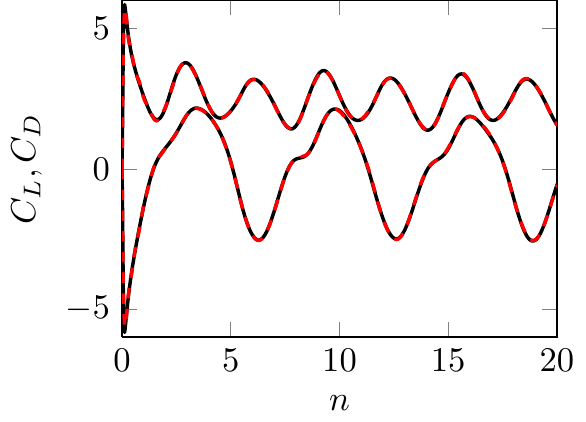}
                \includegraphics{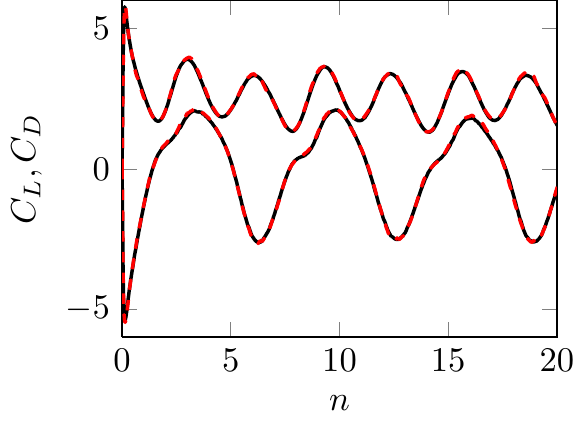}
            }
 \caption{Predicted lift and drag coefficients (bottom and top lines, respectively) for the model FSI problem with a heaving cylindrical body: HDM (solid),
	     ROM (dashed)} \label{fig:Cylinder_C_L C_D}
\end{figure}

Figure~\ref{fig:Cylinder_C_L C_D} displays the evolution of the instantaneous coefficients of lift and drag, $C_L$ and $C_D$,
where $C_D : = F_x /\tfrac{1}{2}\rho u_\infty ^2 d$ and $\rho u_\infty ^2 d=1$. The solid curves correspond to computations
performed using the EBM HDM, whereas the dashed curves correspond to counterparts performed using the
constructed nonlinear ROMs (and a single iteration of the ALS algorithm in the case of the EBM ROMs). The reader can observe that,
as expected, the solutions predicted using both sets of ROMs exhibit comparable accuracies and converge to their counterparts
obtained using the EBM HDM.

\subsection{Two-dimensional turbulent fluid-structure interaction problem in a cavity}

\begin{figure}
    \centering
        \includegraphics{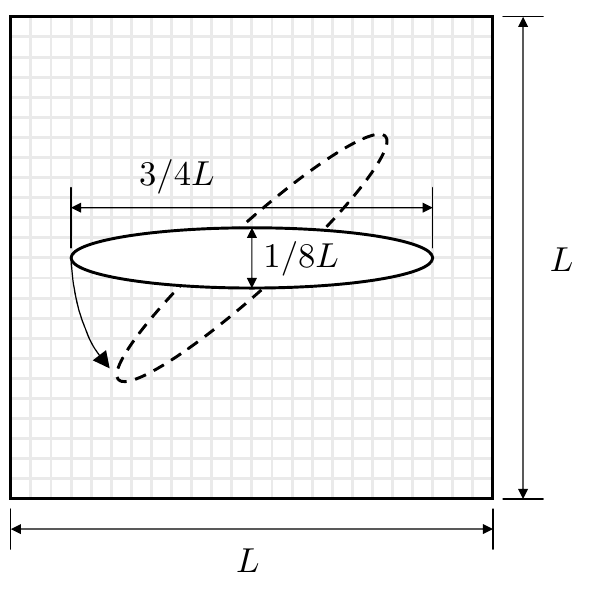}
    \label{fig:Proppeler_geometry}
    \caption{FSI problem with a rotating ellipsoidal body}
\end{figure}

Finally, the two-dimensional computation of a viscous flow inside a square cavity with a rotating ellipsoidal body is considered
here. Specifically, the objective is to predict the flow past the ellipsoidal body during the time-interval in which it
rotates a full $360$ degrees. This FSI problem with large-amplitude displacements and rotations is chosen because unlike the
previous two problems, it cannot be solved using a body-fitted ALE framework without some form of repeated remeshing to avoid
mesh entangling. Hence, it is representative of a large family of FSI problems for which the EBM framework is preferred, if not
essential. The geometry of this problem is illustrated in Figure~\ref{fig:Proppeler_geometry}.

Again, the fluid is modeled as a perfect gas with the ratio of specific heats $\gamma = 1.4$.
The Reynolds number based on the ellipse tip is set to $\operatorname{Re} = 200$ and the Prandtl number to
${\rm Pr} = 0.72$.  The flow is assumed to be initially quiescent.

\begin{figure}
\centering
    \subfigure[EBM HDM\label{sufig:Propeller_DNS}]
    {
        \includegraphics{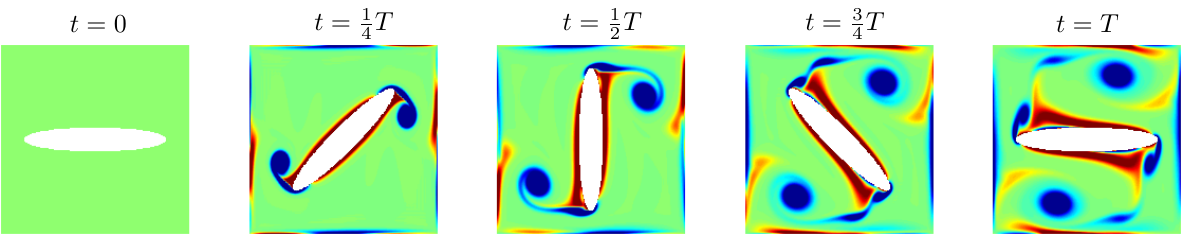}
    }
    \subfigure[EBM ROM with $k=10$ \label{sufig:Propeller_n10}]
    {
        \includegraphics{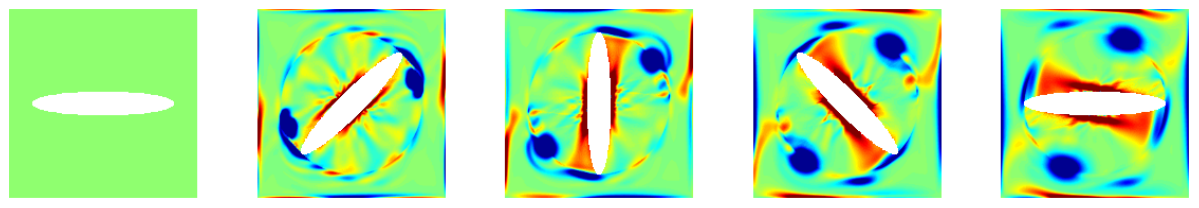}
    }
    \subfigure[EBM ROM with $k=20$ \label{sufig:Propeller_n20}]
    {
        \includegraphics{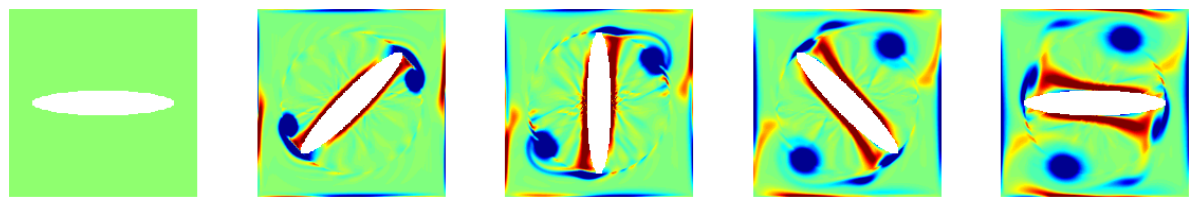}
    }
    \subfigure[EBM ROM with $k=40$ \label{sufig:Propeller_n40}]
    {
        \includegraphics{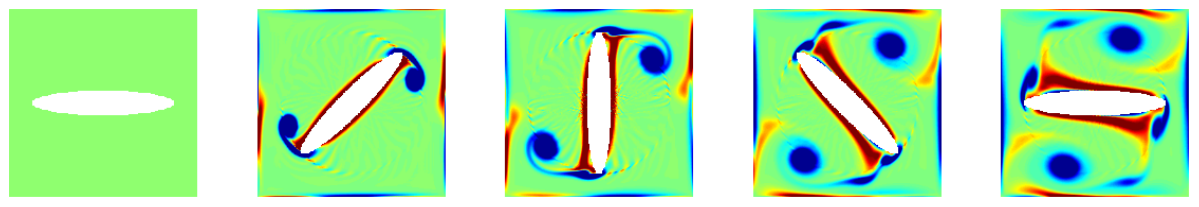}
    }
\caption{Vorticity contours during the rotation of the ellipsoidal body}\label{fig:Propeller}
\end{figure}

The square domain is discretized using a uniform Cartesian grid with $256 \times 256$ elements.
The same EBM outlined for the previous FSI problem with a heaving cylindrical body is applied to its solution.
Computational snapshots of the vorticity contours obtained using the EBM HDM and three different EBM ROMs of dimensions
varying between $k = 10$ and $k = 40$ are illustrated in Figure~\ref{fig:Propeller}. All EBM ROMs are constructed
with a ROB $\bm{U}^{(1)}$ obtained using a single iteration of the ALS algorithm.


As in the previous two cases, the EBM ROM solution is shown to converge to its EBM HDM counterpart
when the dimension $k$ of the ROM is increased. In particular, the reader can observe in Figure~\ref{sufig:Propeller_n20}
that the vorticity field predicted by the EBM ROM of dimension $k=20$ reproduces remarkably well that obtained using the EBM HDM.
For $k=40$, the vorticity fields predicted by the EBM ROM and EBM HDM are almost indistinguishable, thereby demonstrating
the effectiveness of the proposed method for constructing ROBs and ROMs for embedded boundary models.

\section{Conclusions}
\label{sec:CON}

Embedded boundary models are popular for the solution of nonlinear problems with evolving domains and/or moving interfaces.
The application of projection-based model order reduction methods to the construction of reduced-order versions of such models
calls for the construction first of their underlying reduced-order bases. The collection and compression of solution snapshots
using various instances of a computational model of interest is both a popular and effective approach for generating
reduced bases. For a traditional interface-fitted computational model, it is a straightforward procedure that has been popularized
by the Proper Orthogonal Decomposition and Singular Value Decomposition methods. For an embedded boundary computational framework,
this approach faces the problem of dealing with the partitioning of the overall computational domain into a real subdomain and
a ghost one corresponding to the region of the overall computational domain that is occluded by the embedded boundary. For
time-dependent problems, this complication is further exacerbated by the evolution in time of this partitioning as it implies
an evolution in time of the partitioning between meaningful (real) and meaningless (ghost) entries of the
collected solution snapshots. This makes the problem of compressing these snapshots in view of constructing a reduced-order
basis a difficult task. To address this issue, this paper formulates the snapshot compression problem as a weighted low-rank
approximation problem where the binary weighting identifies the evolving component of the individual simulation snapshots,
and proposes to solve this problem using the Alternating Least Squares (ALS) algorithm. This approach is applicable
in principle to any embedded boundary value problem. It is successfully demonstrated in this paper for three different
fluid-structure interaction problems of increasing complexity. In all considered cases, it is shown to deliver reduced-order bases
and models that effectively reproduce the high-dimensional solutions even when the flow is vortex-dominated,
and the immersed body underoges large displacements and rotations.

\clearpage
\appendix

\section{MATLAB implementation of the ALS algorithm}
\label{appA}

The following is a simple MATLAB implementation of Algorithm~\ref{Alg:AP} presented in Section~\ref{sec:ALS}.
The inputs of the function \verb=ALS= are the snapshot matrix \verb=X= where the ghost values are set to zero, a binary
weighting matrix \verb=M=, the desired dimension of the subspace of approximation \verb=k=, and the maximum number of
iterations \verb=it_max=. The outputs of this function are the low-rank approximation matrices \verb=U= and \verb=V=.
\begin{lstlisting}
    function [U,V] = ALS(X,M,k,it_max)
    [cN, K] = size(X);
    [U_star, ~, ~] = svd(X,'econ');
    U = U_star(:,1:k);
    clear U_star;
    V = zeros(k,K);
    for p = 1:it_max
        for j = 1:K
            J = find(s(:,j));
            x = full(X(J,j));
            u = U(J,:);
            V(:,j) = u\x;
        end
        for i = 1:cN
            I = find(M(i,:));
            x = full(X(i,I));
            v = V(:,I);
            U(i,:) = x/v;
        end
    end
\end{lstlisting}

\section{Solution of the weighted low-rank matrix approximation problem when ${\rm rank}(M)=1$ }
\label{appB} 

If $\bm{M}\in \{0,1\}^{cN \times K}$ is a rank 1 weighting matrix, it can be written as $\bm{M}=\bm{s} \bm{t}^{\rm T}$
for some $\bm{s} \in \mathbb{R}^{cN}$ and $\bm{t} \in \mathbb{R}^{K}$. In this case,
\begin{subequations}
\label{eq:RES1}
\begin{align}
\| \bm{M} \odot (\bm{X} - \widetilde{\bm{X}}) \|_F^2 &= \sum\limits_{i,j} {s_i t_j (\bm{X} - \bm{U}\bm{V})_{i,j}^2 }  \nonumber\\
&= \sum\limits_{i,j} {\left( {\sqrt {s_i t_j } X_{i,j}  - \left( {\sqrt {s_i } U_{i,:} } \right)\left( {\sqrt {t_j } V_{:,j} } \right)} \right)} ^2
\end{align}
\end{subequations}
Defining $\bm{X}^{\prime} \in \mathbb{R}^{cN \times K}$ as $X^{\prime}_{i,j} = \sqrt {s_i t_j} X_{i,j}\,\forall i,j$,
$\bm{U}^{\prime} \in \mathbb{R}^{cN\times k}$ as $U^{\prime}_{i,:} = \sqrt {s_i} U_{i,:} \, \forall i$, and
$\bm{V}^{\prime} \in \mathbb{R}^{k\times K}$ as $V^{\prime}_{:,j} = \sqrt {t_j} V_{:,j} \, \forall j$, the above result
can be re-written as
\begin{subequations}
\label{eq:RES2}
\begin{align}
\|\bm{M} \odot (\bm{X} - \widetilde{\bm{X}}) \|_F^2 &=\| (\bm{X}^{\prime} - \widetilde{\bm{X}^{\prime}}) \|_F^2
\end{align}
\end{subequations}
where $\widetilde{\bm{X}^{\prime}} = \bm{U}^{\prime}\bm{V}^{\prime}$. This shows that when $\bm{M}$ is a rank 1 weighting matrix,
the weighted low-rank approximation problem~(\ref{Eqn:WLRA}) reduces to its un-weighted counterpart~(\ref{Eqn:LRA}).

An example of a rank 1 weighting matrix $\bm{M}$ is the weighting matrix appropriate for an embedded boundary model where the
embedded body $B$ is stationary --- that is, $\bm{m}^n=\bm{m}^*, \forall n$ --- but the flow is unsteady. In this case,

\begin{equation}
    \bm{M}: = \left[ {\begin{array}{*{20}c} {\bm{m}^* } & {\bm{m}^* } &  \cdots  & {\bm{m}^* }  \end{array} } \right]
\end{equation}

\section*{Acknowledgments}
The authors acknowledge partial support by the Office of Naval Research under Grant N00014-11-1-0707, and partial support by
the Army Research Laboratory through the Army High Performance Computing Research Center under Cooperative
Agreement W911NF-07-2-0027. The content of this publication does not necessarily reflect the position or policy of any
of the aforementioned institutions, and no official endorsement should be inferred.

\section*{References}
\bibliographystyle{elsarticle-num}

\end{document}